%% file: ms.tex
\documentclass{nature}
\usepackage{url}
\usepackage[colorinlistoftodos]{todonotes}
\usepackage{times}
\usepackage{gensymb}
\usepackage{lineno}
\usepackage[T1]{fontenc}
\usepackage{color}
\usepackage{multirow}
\usepackage{setspace}
\usepackage{array}
\usepackage{makecell}
\usepackage{tikz}
\usepackage{pifont}

\title{\large{Identifying Pediatric Vascular Anomalies With Deep Learning}}

\author
{Justin Chan,$^{1}\star$ Sharat Raju$^{2,3}\star$, Randall Bly$^{2,3\ast}$, Jonathan A. Perkins$^{2,3\ast}$ and Shyamnath Gollakota$^{1\ast}$\\
\normalsize{$^{1}$Paul G. Allen School of Computer Science and Engineering, University of Washington, WA}\\
\normalsize{$^{2}$Department of Otolaryngology -- Head and Neck Surgery, University of Washington, WA}\\
\normalsize{$^{3}$Seattle Children's Hospital and Research Institute, Seattle, WA}\\
\normalsize{$^\star$Equal contribution first authors}\\
\normalsize{$^\ast$Correspondence authors: gshyam@uw.edu, jonathan.perkins@seattlechildrens.org, randbly@uw.edu}
\\
}

\begin{document}

\maketitle

\begin{abstract}
\input{abs-4.tex}
\end{abstract}
\vskip 0.4in
\input{intro-5.tex}
\input{results-2.tex}

\input{discussion-4.tex}

\input{methods-1}

\section*{Supplementary Materials}
Supplementary Fig. 1. t-SNE visualization of the CNN's weights.\\
Supplementary Fig. 2. Individual confusion matrices for pediatricians taking the survey.\\
Supplementary Fig. 3. User interface of CNN running on a smartphone.\\
Supplementary Table 1. Cross-validation performance over six classes.\\
Supplementary Table 2. $F_1$ score on test set of six classes.

\input{acks-1}
\section*{References}
\bibliography{ms}
\bibliographystyle{naturemag}

\clearpage
\pagebreak


\begin{figure}
    \centering
    \includegraphics[width=.8\textwidth]{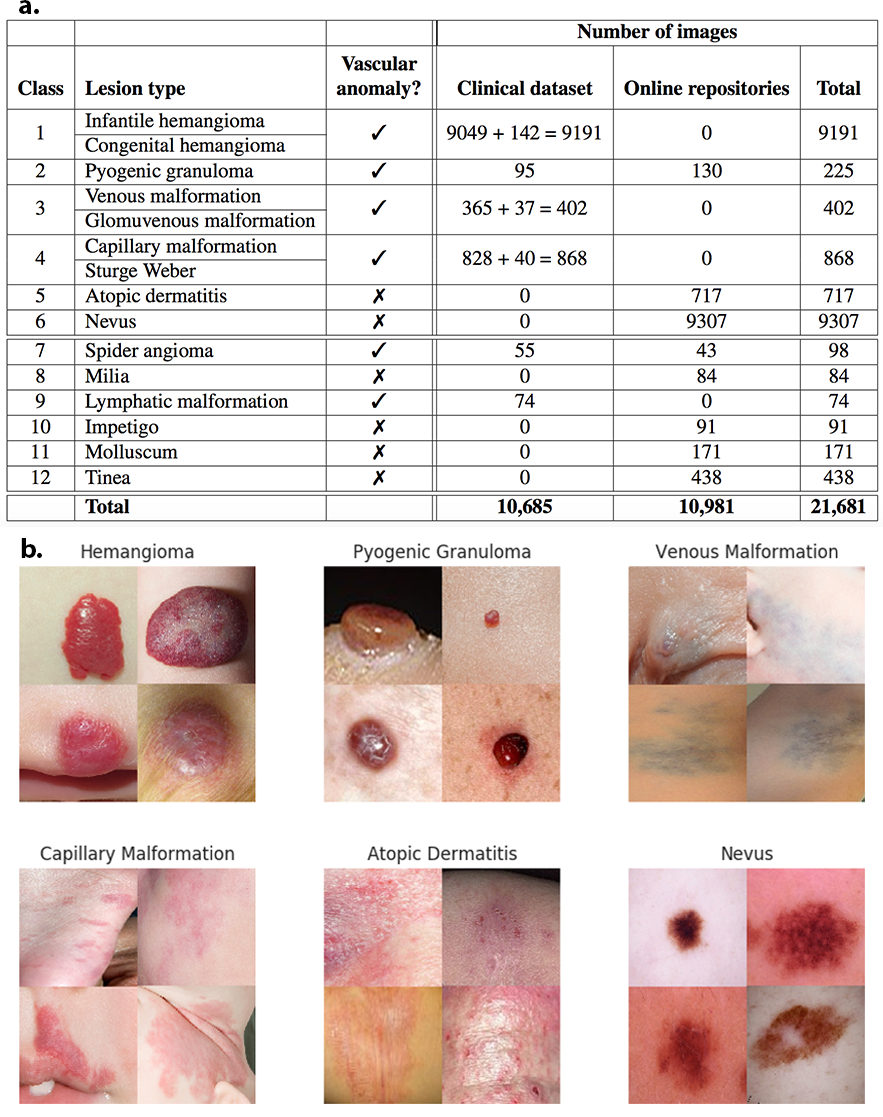}
    \vspace{-1em}
    \caption{{\bf Pediatric skin lesions taxonomy and example images.} {\bf a.} Taxonomy of 12 pediatric skin lesions. The table shows the composition of images which were obtained from the Seattle Children's clinical dataset and dermatologist-curated online repositories. The upper half of the table represents a 6-class subset of our taxonomy where images in each class are comparatively data abundant and consist of pediatric skin lesions more commonly seen in practice. {\bf b.} Example images of the lesions in the 6-class subset of our taxonomy. The images demonstrate the visual similarity between different lesion types.}
    \label{fig:taxonomy}
\end{figure}

\clearpage
\pagebreak

\clearpage
\pagebreak

\begin{figure}
    \centering
    \includegraphics[width=\textwidth]{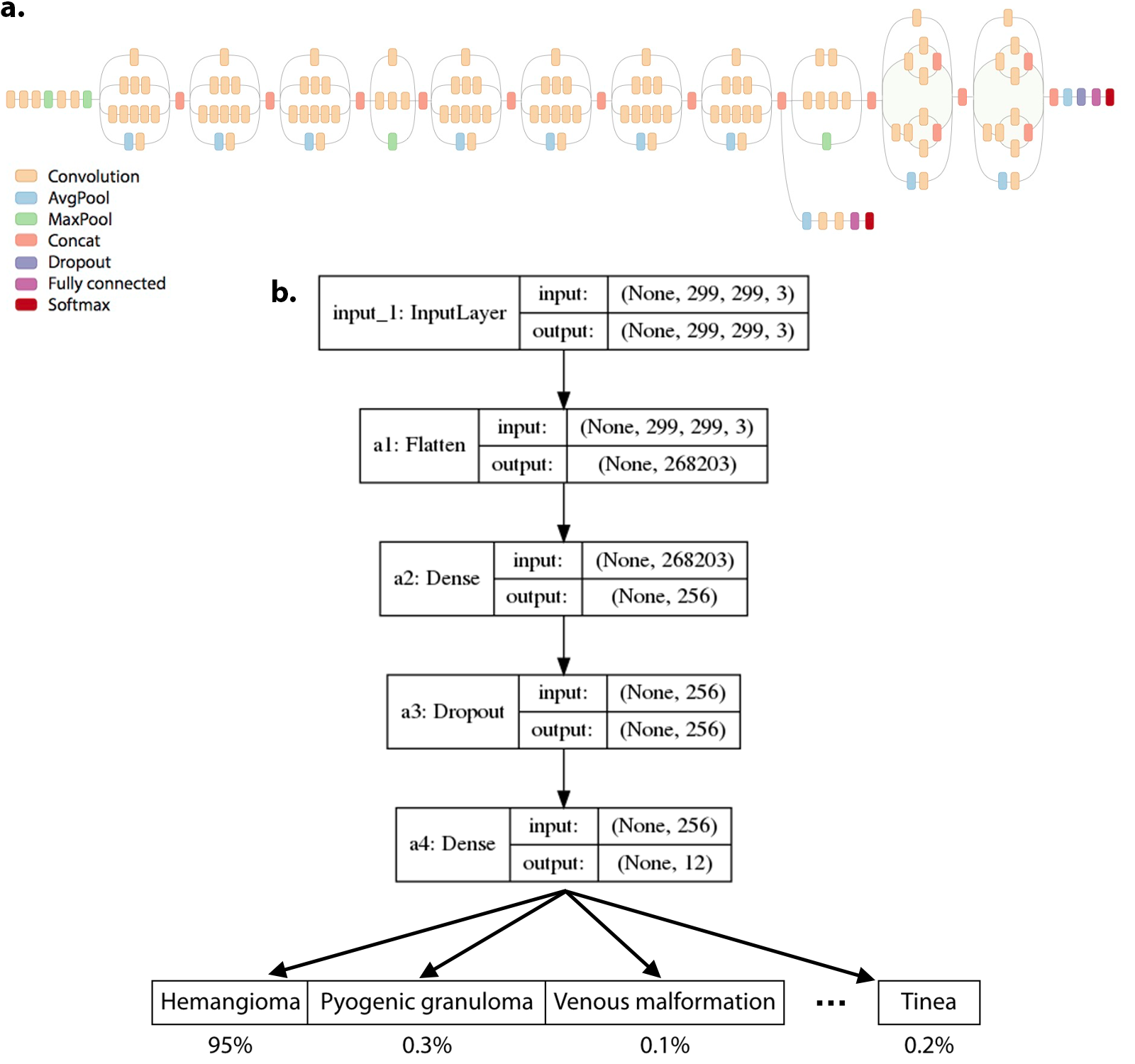}
    \caption{{\bf Convolutional neural network architecture.} {\bf a.} We leverage the InceptionV3 network architecture that has been pretrained on the ImageNet dataset. {\bf b.} The weights of the network are fine-tuned on a fully-connected layer of 256 nodes, and outputs a vector of probabilities indicating prediction likelihood for each of the 12 pediatric skin lesion classes.}
    \label{fig:arch}
\end{figure}

\clearpage
\pagebreak

\begin{table}
\centering
\begin{tabular}{ |l||c|c|  }
 \hline
 {\bf Lesion type} & {\bf AUC (95\% CI)} & {\bf $F_1$ score}\\
 \hline
Hemangioma & 0.9608 (0.9514 -- 0.9701) & 0.9188 (0.8988 -- 0.9388) \\ \hline
Pyogenic granuloma & 0.9875 (0.9735 -- 1.0015) & 0.9696 (0.9571 -- 0.9822) \\ \hline
Venous+gelomuvenous malformation & 0.9750 (0.9684 -- 0.9816) & 0.9374 (0.9237 -- 0.9511) \\ \hline
Capillary malformation+Sturge Weber & 0.9762 (0.9714 -- 0.9810) & 0.9373 (0.9228 -- 0.9519) \\ \hline
Atopic dermatitis & 0.9558 (0.9481 -- 0.9634) & 0.8996 (0.8783 -- 0.9210) \\ \hline
Nevus & 0.9994 (0.9990 -- 0.9998) & 0.9894 (0.9850 -- 0.9938) \\ \hline
Spider angioma & 0.9716 (0.9603 -- 0.9829) & 0.9354 (0.9165 -- 0.9543) \\ \hline
Lymphatic malformation & 0.9279 (0.8957 -- 0.9602) & 0.8899 (0.8668 -- 0.9129) \\ \hline
Milia & 0.9967 (0.9957 -- 0.9977) & 0.9751 (0.9688 -- 0.9815) \\ \hline
Impetigo & 0.9825 (0.9797 -- 0.9853) & 0.9372 (0.9298 -- 0.9446) \\ \hline
Molluscum & 0.9863 (0.9845 -- 0.9881) & 0.9524 (0.9429 -- 0.9619) \\ \hline
Tinea & 0.9573 (0.9498 -- 0.9648) & 0.8984 (0.8875 -- 0.9093) \\ \hline \hline
{\bf Average}&0.9731&0.9367 \\ \hline
\end{tabular}
\caption{{\bf Cross-validation performance of classifier over 12 classes.} The table shows the AUC and weighted $F_1$ score of the classifier for each class after performing ten-fold cross-validation over the full taxonomy of 12 lesion types. Additionally, we show the averaged AUC and $F_1$ score.}
\label{tab:crossval12}
\end{table}

\clearpage
\pagebreak

\begin{figure}
    \centering
    \includegraphics[width=\textwidth]{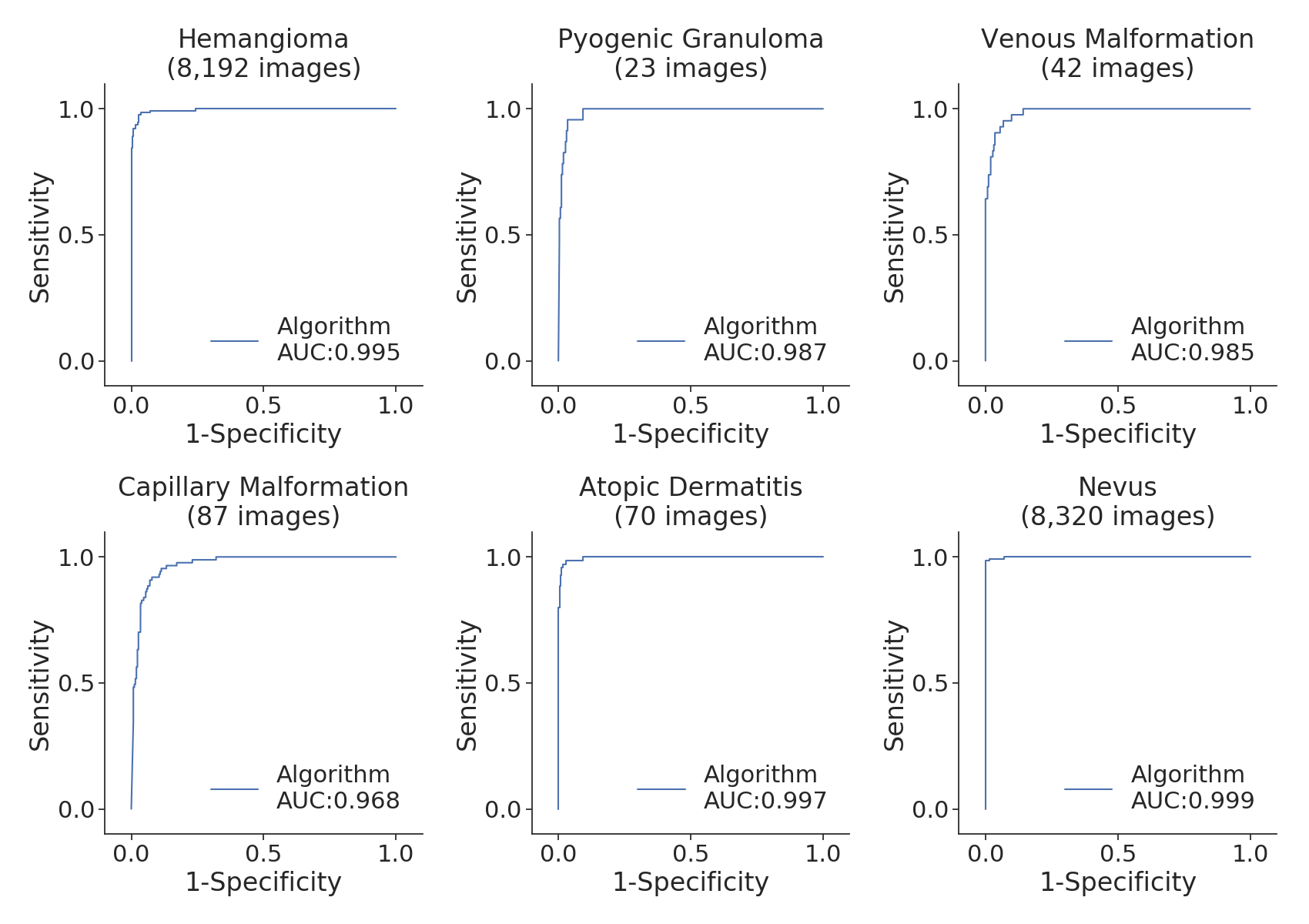}
    \caption{{\bf Classifier performance on a previously unseen test set.} The receiver-operating curves show the performance of our classifier on a previously unseen test set of images from our 6-class taxonomy.}
    \label{fig:roc}
\end{figure}

\clearpage
\pagebreak

\begin{figure}
    \centering
    \includegraphics[width=\textwidth]{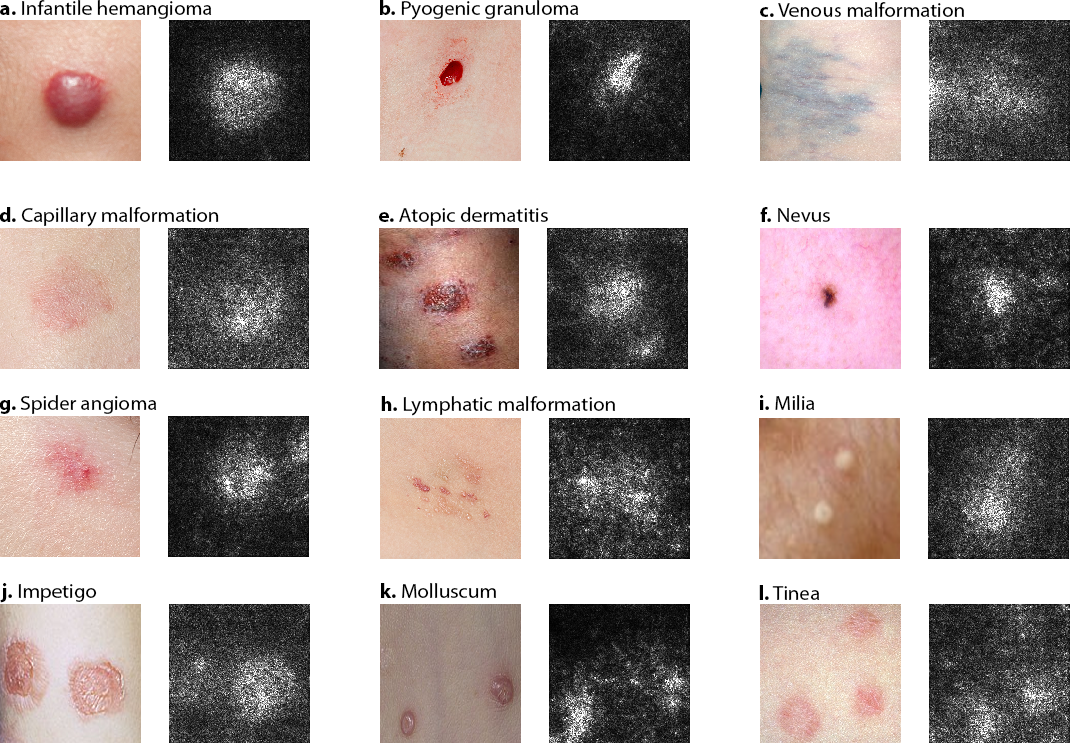}
    \caption{{\bf Saliency maps for 12 pediatric skin lesion classes.} The saliency maps show which pixels within the image contribute most to the classifier's predictions. White pixels show a greater contribution to the prediction. Each image shows either one or more distinct lesions. The corresponding saliency maps show a correlation between the location of the lesion and the pixels used by the classifier to come to a prediction.
    {\bf a.} Infantile hemangioma 
    {\bf b.} Pyogenic granuloma
    {\bf c.} Venous malformation 
    {\bf d.} Capillary malformation 
    {\bf e.} Atopic dermatitis
    {\bf f.} Nevus
    {\bf g.} Spider angioma
    {\bf h.} Lymphatic malformation 
    {\bf i.} Milia  
    {\bf j.} Impetigo
    {\bf k.} Molluscum
    {\bf l.} Tinea
    }
    \label{fig:saliency}
\end{figure}


\clearpage
\pagebreak

\begin{figure}
    \centering
    \includegraphics[width=\textwidth]{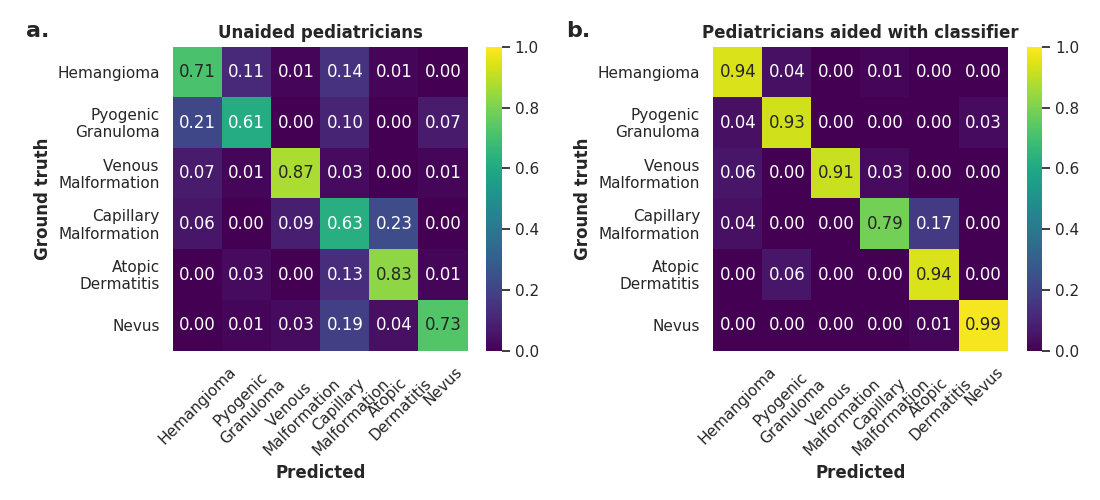}
    \includegraphics[width=.5\textwidth]{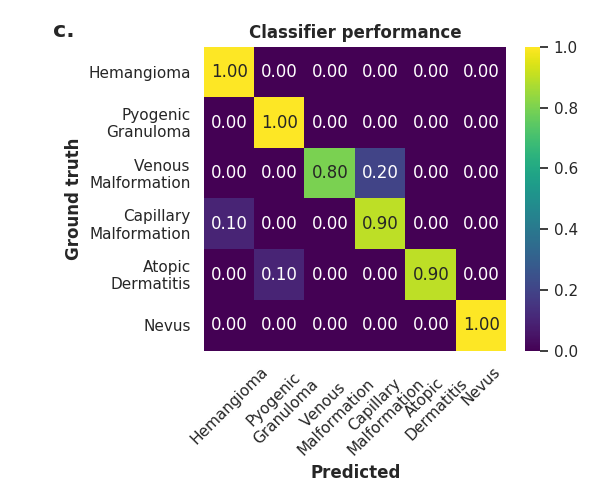}
    \caption{{\bf Diagnostic performance of pediatricians without and with the aid of the classifier.} The confusion matrices show the diagnostic accuracies of seven pediatricians in a survey containing 60 images drawn uniformly from six pediatric skin lesions. The figure shows the performance {\bf a.} without and {\bf b.} with the aid of our classifier and {\bf c.} the performance of the classifier itself. With the aid of the classifier, diagnostic accuracies are higher for each of the six classes. When aided with the classifier, the pediatricians achieve higher accuracies for venous malformations and atopic dermatitis compared to the classifier  alone.}
    \label{fig:survey}
\end{figure}

\clearpage
\pagebreak

\singlespacing
\setcounter{figure}{0}
\setcounter{table}{0}
\resetlinenumber
\setcounter{page}{1}
\renewcommand{\figurename}{Supplementary Figure}
\renewcommand{\tablename}{Supplementary Table}

\begin{figure}
    \centering
    \includegraphics[width=\textwidth]{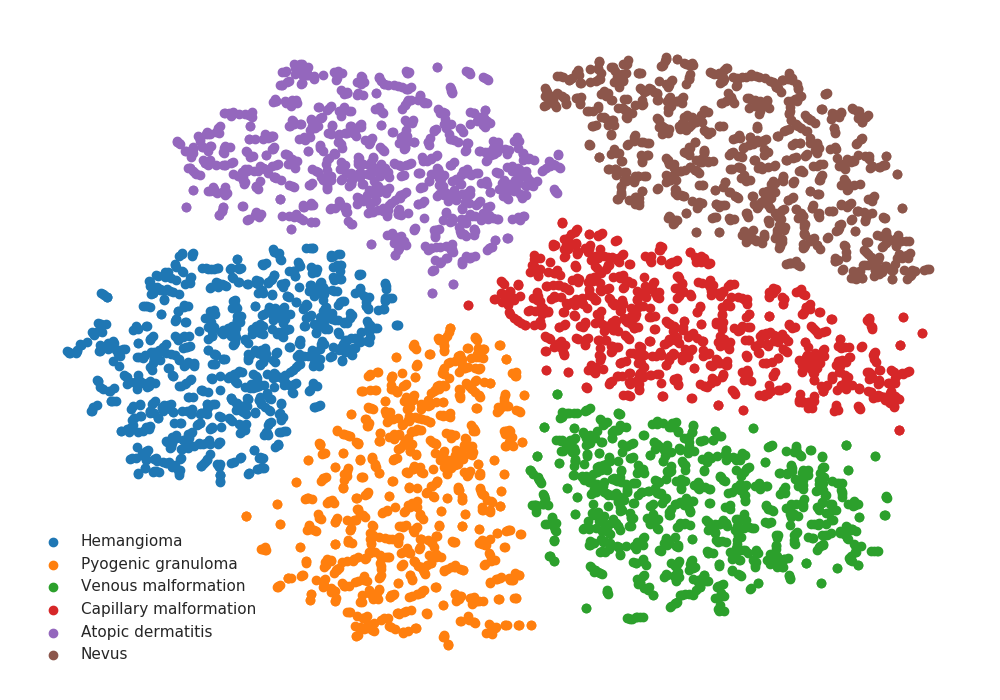}
    \caption{{\bf t-SNE visualization of the CNN's weights.} The visualization is a projection of the 256 features learned at the final layer of our CNN after it is fine-tuned on our dataset of pediatric skin lesions. The visualization is of a random subset of 5,381 images used to train our CNN.}
    \label{fig:tsne}
\end{figure}

\clearpage
\pagebreak

\begin{figure}
    \centering
    \includegraphics[width=.8\textwidth]{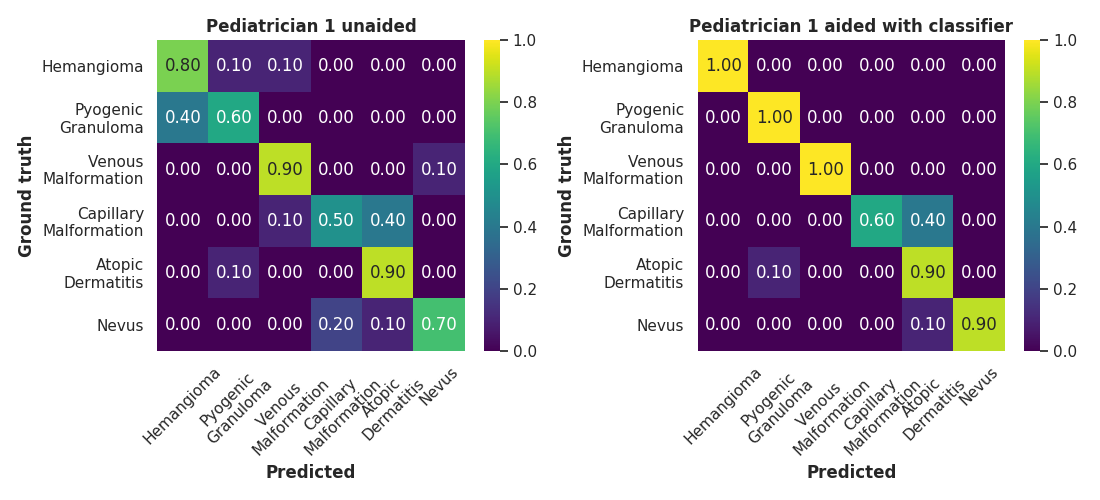}
    \includegraphics[width=.8\textwidth]{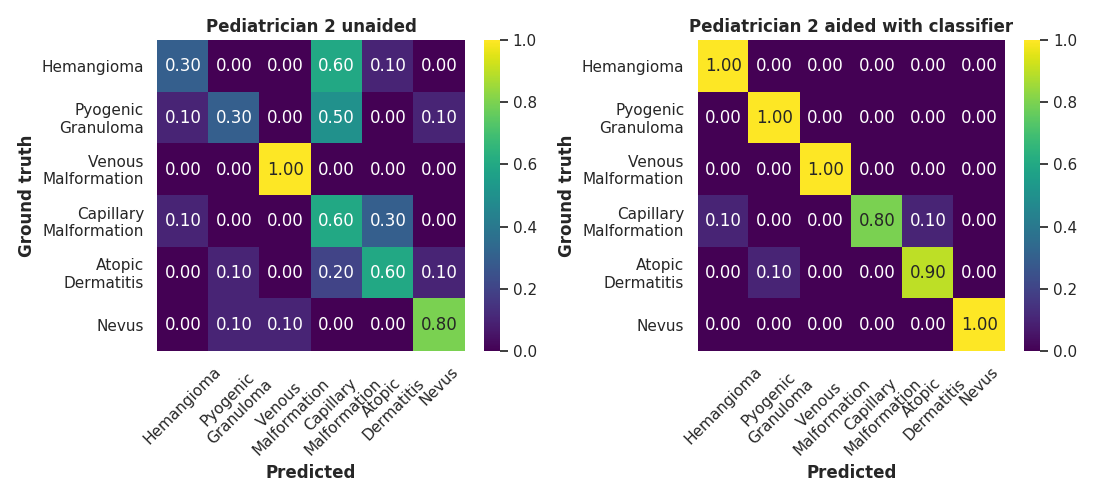}
    \includegraphics[width=.8\textwidth]{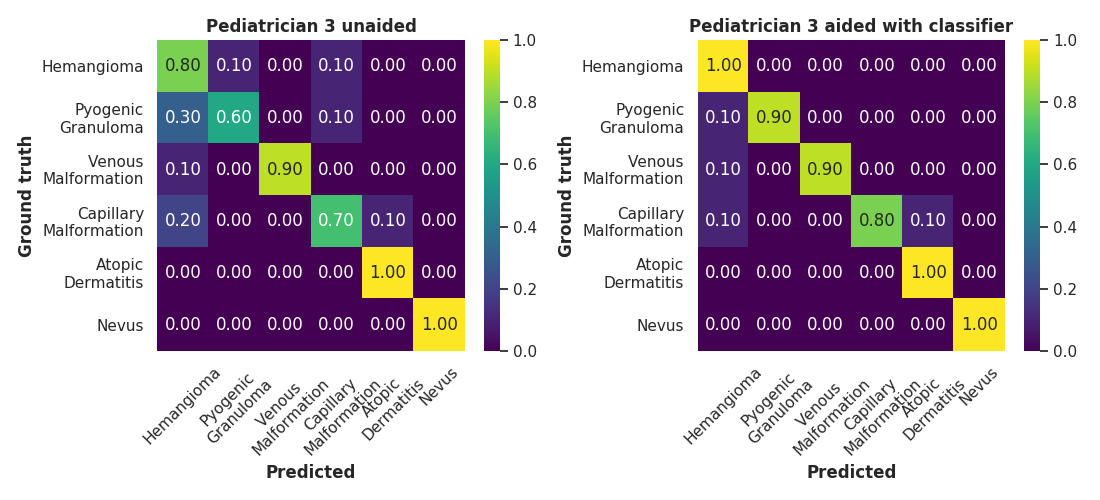}
    \caption{{\bf Individual confusion matrices for pediatricians taking the survey.} Each row is the performance of a single pediatrician. Left and right columns indicate performance when pediatrician is unaided and aided with the classifier's predictions respectively.}
    \label{fig:iconf}
\end{figure}

\clearpage
\pagebreak

\centering
\includegraphics[width=.8\textwidth]{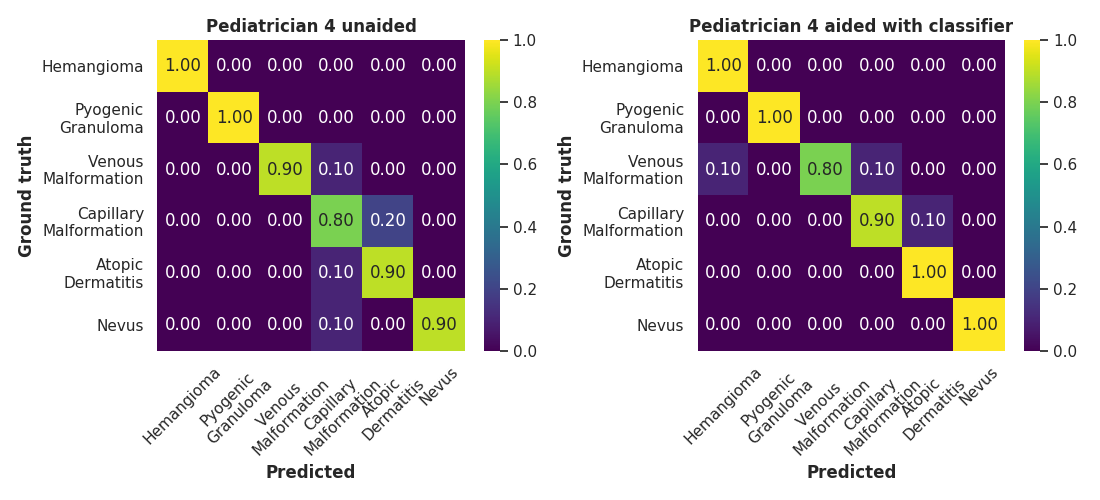}\\
\includegraphics[width=.8\textwidth]{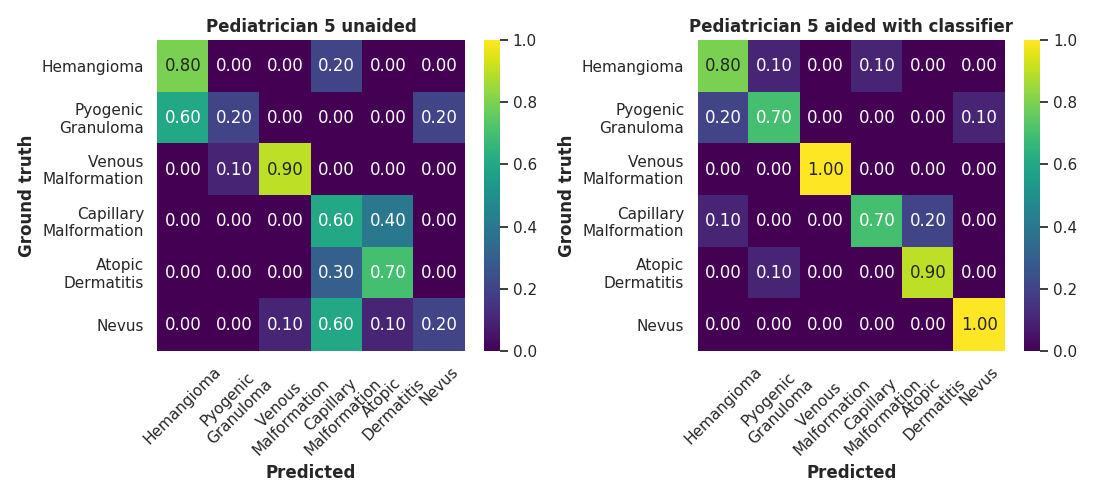}\\
\includegraphics[width=.8\textwidth]{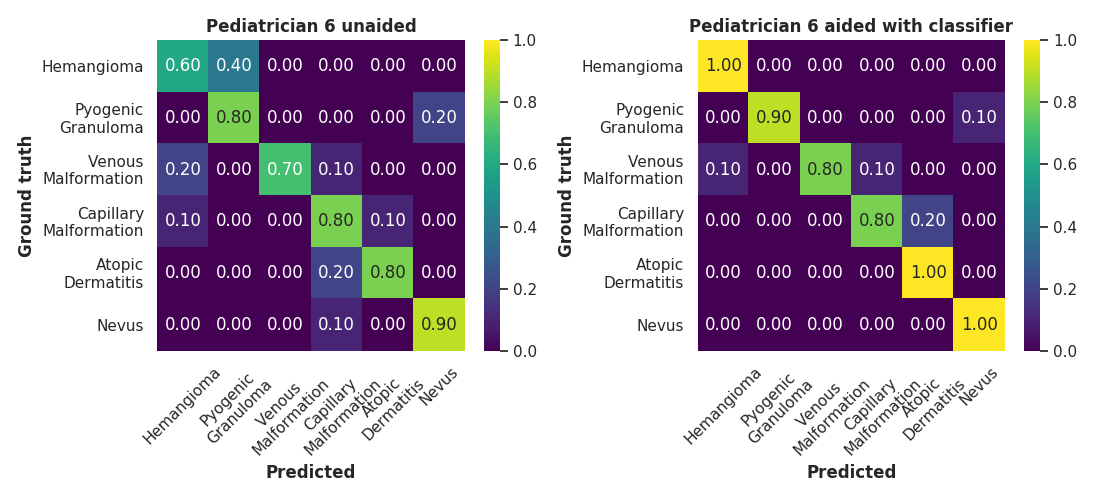}\\
Supplementary Figure~\ref{fig:iconf} continued.

\clearpage
\pagebreak

\centering
\includegraphics[width=.8\textwidth]{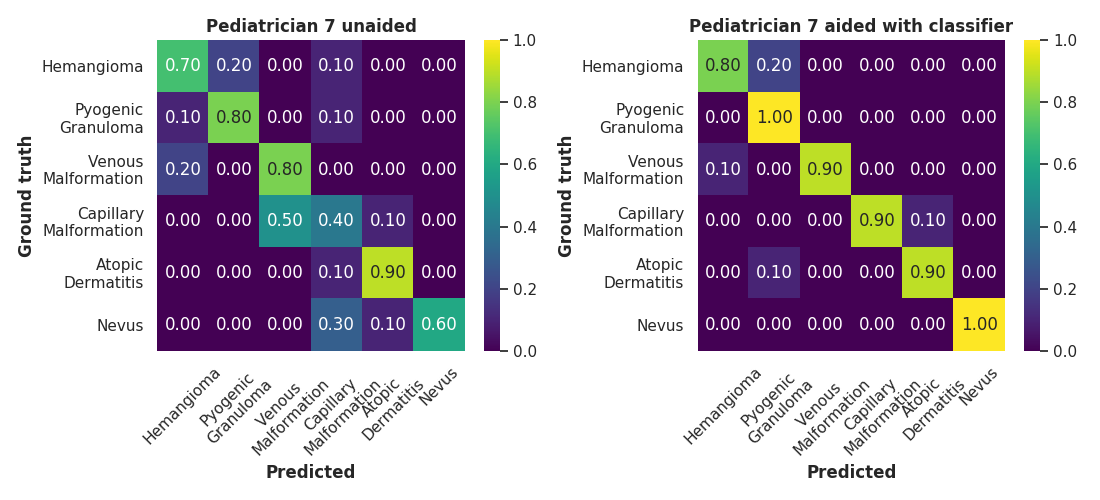}\\
Supplementary Figure~\ref{fig:iconf} continued.

\clearpage
\pagebreak

\begin{figure}
    \centering
    \includegraphics[width=.9\textwidth]{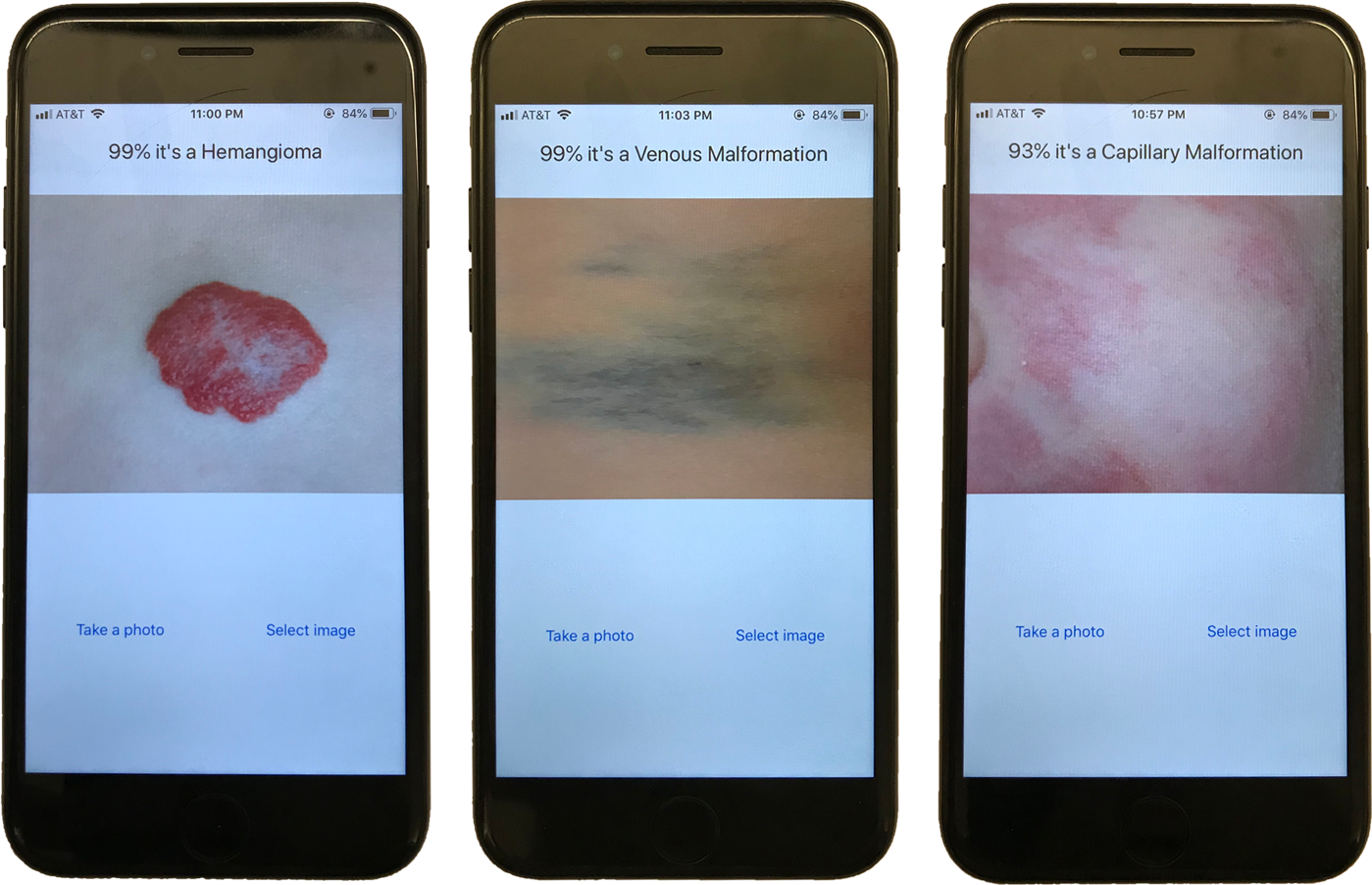}
    \caption{{\bf User interface of CNN running on a smartphone.} The interface allows users to either take a photo or upload an existing image of a pediatric skin lesion. The photo is then passed to our CNN which is running in real time on an iPhone 7. The user is presented with the classifier's prediction and probability output. We show an example of a hemangioma, venous malformation and capillary malformation being classified by our CNN.}
    \label{fig:smartphones}
\end{figure}

\clearpage
\pagebreak

\begin{table}
\centering
\begin{tabular}{ |l||c|c|  }
\hline
{\bf Lesion type} & {\bf AUC (95\% CI)} & {\bf $F_1$ score}\\
\hline
Hemangioma & 0.9621 (0.9546 -- 0.9696) & 0.9084 (0.8943 -- 0.9224) \\ \hline
Pyogenic granuloma & 0.9933 (0.9891 -- 0.9974) & 0.9677 (0.9519 -- 0.9835) \\ \hline
Venous+gelomuvenous malformation & 0.9803 (0.9749 -- 0.9858) & 0.9333 (0.9222 -- 0.9444) \\ \hline
Capillary malformation+Sturge Weber & 0.9729 (0.9646 -- 0.9812) & 0.9232 (0.9099 -- 0.9364) \\ \hline
Atopic dermatitis & 0.9947 (0.9929 -- 0.9964) & 0.9666 (0.9589 -- 0.9743) \\ \hline
Nevus & 0.9997 (0.9996 -- 0.9999) & 0.9916 (0.9891 -- 0.9942) \\ \hline \hline
{\bf Average}&0.98384&0.9485\\ \hline
\end{tabular}
\caption{{\bf Cross-validation performance over six classes.} The table shows the AUC and weighted $F_1$ score obtained for each class when performing ten-fold cross-validation over the 6-class subset of our taxonomy. The table also shows the average AUC and $F_1$ score.}
\label{tab:crossval6}
\end{table}

\clearpage
\pagebreak

\begin{table}
\centering
\begin{tabular}{ |l||c|  }
\hline
{\bf Lesion type} & {\bf $F_1$ score}\\
\hline
Hemangioma & 0.9700 \\ \hline
Pyogenic granuloma & 0.9810 \\ \hline
Venous+gelomuvenous malformation & 0.9627 \\ \hline
Capillary malformation+Sturge Weber & 0.9548 \\ \hline
Atopic dermatitis & 0.9824 \\ \hline
Nevus & 0.9883 \\ \hline \hline
{\bf Average}&0.9732\\ \hline
\end{tabular}
\caption{{\bf $F_1$ score on test set of six classes.} The table shows the weighted $F_1$ score when the classifier is evaluated on the test set of six classes. The probability threshold is set to maximize the sensitivity and specificity of that particular class.}
\label{tab:test6}
\end{table}

\end{document}

%% file: abs-4.tex
Vascular anomalies, more colloquially known as birthmarks, affect up to 1 in 10 infants. Though many of these lesions self-resolve, some types can result in medical complications or disfigurement without proper diagnosis or management. Accurately diagnosing vascular anomalies is challenging for pediatricians and primary care physicians due to subtle visual differences and similarity to other pediatric dermatologic conditions. This can result in delayed or incorrect referrals for treatment. To address this problem, we developed a convolutional neural network (CNN) to automatically classify images of vascular anomalies and other pediatric skin conditions to aid physicians with diagnosis. We constructed a dataset of 21,681 clinical images, including data collected between 2002--2018 at Seattle Children's hospital as well as five dermatologist-curated online repositories, and built a taxonomy over  vascular anomalies and other common pediatric skin lesions. The CNN achieved an average AUC of 0.9731 when ten-fold cross-validation was performed across a taxonomy of 12 classes. The classifier's average AUC and weighted $F_1$ score was 0.9889 and 0.9732 respectively when evaluated on a previously unseen test set of six of these classes. Further, when used as an aid by pediatricians (n = 7), the classifier increased their average visual diagnostic accuracy from 73.10\% to 91.67\%. The classifier runs in real-time on a smartphone and has the potential to improve diagnosis of these conditions, particularly in resource-limited areas.

%% file: intro-5.tex
\section*{Introduction}
Vascular anomalies, colloquially known as ``birthmarks'', affect between 5 and 13\% \cite{nosher2014vascular,greene2008risk} of all infants. Most are associated with changes in the appearance of the overlying skin, and initial diagnosis is commonly made by history and physical exam, a large part of which is visual inspection. 

Vascular anomalies encompass a large number of diagnoses which are classified by the International Society for the Study of Vascular Anomalies (ISSVA) into vascular tumors and vascular malformations.  Some vascular anomalies, such as certain infantile hemangiomas, will fade with time and can be managed medically or with observation in the primary care setting \cite{juern2010nevus,haggstrom2006prospective}. In contrast, venous malformations, congenital hemangiomas, and lymphatic malformations are best managed early by a multidisciplinary team of specialists including dermatologists, surgeons, and  pediatricians. Without proper treatment, bleeding, infection, permanent disfigurement, or airway complications may occur. As a result, early and correct diagnosis is critical to prevent delays in management \cite{lee2018vascular,greene2011vascular}.

Accurate diagnosis of vascular anomalies is challenging. Vascular anomalies can occur on any surface of the body, and the skin manifestations can include a wide range of sizes and hues. As a result, these lesions can be difficult to visually differentiate, and it has been shown that as low as 31 -- 53\% of vascular anomalies have a correct diagnosis at the time of referral \cite{greene2011vascular,levin2013room,macfie2008diagnosis}. Additionally, some of these anomalies may be confused with common pediatric dermatologic conditions. This can lead to misplaced expectations from patients and families, and in some cases, delays in delivery of care. Thus, to assist pediatricians and other primary care physicians with accurate diagnosis, we develop an image classification system which uses a convolutional neural network (CNN) to automatically classify images of vascular anomalies. 

A computer-aided system for classifying vascular anomalies currently does not exist. At most tertiary vascular anomaly centers, proper diagnosis and staging relies on a combination of modalities including clinical history and exam, imaging, angiography, tissue biopsy, and multidisciplinary consensus \cite{greene2011vascular,nosher2014vascular}. However, these are not readily available in the primary care setting.

In this study, we construct a dataset comprising 21,681 labeled images of cutaneous skin lesions spanning 15 different pediatric dermatologic conditions including nine vascular anomalies. 10,700 images in the dataset were collected from Seattle Children's Hospital during 2002--2018 as clinical photographs were routinely obtained for all patients visiting the vascular anomaly clinic. Additionally, we include 10,981  images from five dermatologist-curated online repositories to supplement image classes that were sparse or did not exist in our clinical dataset. We developed a CNN,\cite{krizhevsky2012imagenet} a type of deep learning system optimized for image classification, to visually identify and diagnose vascular anomalies. CNNs are able to automatically learn representations of input data and make predictions without the need for extensive pre-processing and feature engineering\cite{nixon2012feature}. CNNs have shown specialist-level accuracy at diagnosing diseases such as melanoma\cite{esteva2017dermatologist}, pneumonia\cite{rajpurkar2017chexnet}, diabetic retinopathy\cite{gulshan2016development} and cardiovascular risk\cite{poplin2018prediction}. We demonstrate that such a tool can 
{\it improve} diagnostic accuracy {for vascular anomalies and other pediatric dermatologic conditions} among a cohort of pediatricians.


%% file: results-2.tex
\section*{Results}
All images collected from Seattle Children's Hospital were diagnosed by biopsy, computerized tomography (CT), angiography, ultrasonography, or specialist consensus when possible. {Demographic information was collected for images associated with a valid medical record number and date of photography. The female to male ratio was 2.01 and the median age was 4 (inter-quartile range: 2) months.} The images selected for inclusion were further curated by three vascular anomaly surgeons to exclude lesions without cutaneous manifestation. All images were de-identified and cropped to only include the area relevant for diagnosis. These images are organized into a taxonomy of 12 different vascular anomaly and pediatric dermatologic classes as shown in Fig.~\ref{fig:taxonomy}a. Vascular anomalies classified under the same taxonomy, such as venous and glomuvenous malformations, 
were grouped together. Fig.~\ref{fig:taxonomy}b shows example images for six of these classes, illustrating the visual similarity among these lesions.

We use a technique known as transfer learning\cite{pan2010survey} to classify the images of our dataset. We leverage the InceptionV3 CNN\cite{szegedy2016rethinking} that has been pre-trained on the 2012 ImageNet Large Scale Visual Recognition Challenge (ILSVRC)\cite{russakovsky2015imagenet} containing 1000 classes, and then fine-tune the weights of the network to our dataset of images. To do this we remove the final softmax layer that produces outputs for the 1000 ImageNet classes. We then add our own layers as shown in Fig.~\ref{fig:arch} that ends in a softmax output that produces probability outputs for our 12 image classes. 

We first validate our classifier using ten-fold cross-validation\cite{kohavi1995study}. Due to an unequal number of images in each skin lesion class, images within the dataset are augmented using label-preserving transformations\cite{cirecsan2012multi,simard2003best}. We augment each class used for cross-validation to 1000 images. This is to ensure the classifier is not overly biased towards one particular class, or fails to learn a sparse class. This class size was chosen to align with the ImageNet dataset\cite{deng2009imagenet}, which contains an average of 500--1000 images within each subcategory. Specifically, images are rotated at a random angle and a horizontal and vertical flip are also applied at random. Shear up to an intensity of 0.2 and a zoom in the range 0.8 and 1.2 is applied to the image\cite{esteva2017dermatologist,krizhevsky2012imagenet}. Classes that had more than 1000 images were randomly sampled for cross-validation. Images are then resized to 299$\times$299 dimensions to be compatible with the input dimensions of our pre-trained CNN. The data split between the training and validation sets was such that images of the same lesion from multiple angles did not exist in both the training and validation set for any fold. This ensures that the CNN does not leverage patient-specific information when making a prediction. The cross-validation results in Table~\ref{tab:crossval12} show the AUCs, confidence intervals and $F_1$ score for each of the 12 classes. To calculate the $F_1$ score, the probability threshold for each class is set to maximize the sum of the sensitivity and specificity on that class' ROC curve. The $F_1$ score is then weighted in accordance to the number of positive and negative examples of that class. The average AUC and $F_1$ score across all classes is 0.9731 and 0.9367 respectively.

Next, we evaluate the test performance of our classifier on a held-out independent unseen dataset. The criteria for a sufficient number of images in a test class is based on the composition of images in the ILSVRC which has 50--100 images per test class. To obtain meaningful test performance values, we selected the first six classes in Fig.~\ref{fig:taxonomy}a that were sufficiently data abundant and most commonly seen in practice based on consensus by vascular anomaly specialists and dermatologists. For classes with more than 1000 images, images not sampled for cross-validation are used in the test set. For the remaining classes, 10\% of the total number of images in that class were withheld from the prior cross-validation step, and included in the test set. {Using the same CNN architecture as before, we train and evaluate a classifier over these six classes.} The cross-validation results over these six classes are shown in Supplementary Table~\ref{tab:crossval6}, the average AUC and $F_1$ score across six classes is 0.98384 and 0.9485 respectively. We show the individual receiver-operating characteristic (ROC) curves for the classifier's performance on previously unseen test data from each of these six classes in Fig.~\ref{fig:survey}. {The average AUC across these six classes is 0.9889 and the average weighted $F_1$ score is 0.9732  (Supplementary Table~\ref{tab:test6}).}

We generate saliency maps in Fig.~\ref{fig:saliency} for an example image in each class of our 12-class taxonomy using integrated gradients\cite{sundararajan2017axiomatic}. {The maps confirm that the CNN places more weight on the pixels representing the lesion compared to surrounding skin when making a prediction.} Additionally, we visualize the features learned at the last layer of our CNN classifier using t-SNE for our 6-class taxonomy in Supplementary Fig.~\ref{fig:tsne}. The projection of features onto a 2-D space shows that each class is clustered tightly and are separated from the clusters of other classes. 

We next evaluate if our CNN trained on six lesion classes can be used to aid pediatricians to more accurately diagnose vascular anomalies, and thus make appropriate referrals. We presented 60 images from our test set to seven pediatricians. Only clear and visible images were selected for inclusion in this subset. On this subset of 60 test images, our classifier had an accuracy of 93.33\%. Each pediatrician was asked to classify each image into one of six classes. Pediatricians achieved an average accuracy of 73.10\% on this task. The pediatricians were not informed of their accuracy or of the correct labels of the images during their initial pass. They were then presented with the same set of images in a different random order, each annotated with the classifier's predictions, and asked to classify each image. When aided with the classifier's predictions, the average accuracy increased to 91.67\%. We compare the confusion matrix across all pediatricians when they are unaided and aided with our classifier (Fig.~\ref{fig:survey}, Supplementary Fig.~\ref{fig:iconf}). The figure shows that pediatrician accuracy is increased across all six classes. Additionally, when aided with the classifier, pediatricians are able to achieve higher accuracies for venous malformations and atopic dermatitis than when using the classifier alone. Specifically, 5 out of 7 pediatricians classified venous malformations with a higher accuracy than the classifier, achieving 96\% average accuracy compared to the classifier's 80\%. The remaining 2 pediatricians were on par with the classifier. 3 out of 7 pediatricians classified atopic dermatitis more accurately than the classifier, obtaining an average accuracy of 100\% compared to the classifier's accuracy of 90\%; the remaining 4 pediatricians matched the accuracy of the classifier. This suggests that for these classes, combining the computer-aided system with pediatrician expertise can potentially have an advantage over either method alone.

Finally, we evaluate if our classifier can be deployed and executed on a smartphone in real-time (Supplementary Fig.~\ref{fig:smartphones}). On an iPhone 7, the classifier makes real-time predictions within 32~ms.

%% file: discussion-4.tex
\section*{Discussion}
Early and accurate diagnosis of vascular anomalies is essential to minimize complications and ensure appropriate treatment. For many primary care physicians, diagnosis relies on identifying subtle visual clues. We present data that a CNN trained on images of vascular anomalies and other common pediatric skin lesions can enhance the diagnostic accuracy of physicians, which may improve outcomes and optimize referral patterns. 

A limitation of our dataset is the imbalance of images across different classes. This is reflected in the increased prevalence of certain vascular anomalies such as infantile hemangiomas. As a result the number of images available in the test set for evaluation are relatively small for pyogenic granulomas (lobular hemangioma) and venous malformations. Acquiring images of more sparse diagnostic classes would allow us to more accurately estimate the real world performance of the classifier on these lesion types. We note however that a clinical deployment of this system may benefit from intentionally biasing the classifier to incorporate real world prevalence rates of different vascular anomalies when making a prediction. Evaluating the classifier in a larger prospectively obtained cohort may provide a more robust estimate of accuracy and potential clinical impact. Additionally, though the CNN can be deployed on a commodity smartphone, variations in the quality and setting of the photos may affect real-world classification accuracy. In particular, evaluating the classifier on images taken with multiple smartphone cameras would be needed to test how well our classifier generalizes to images obtained in primary care clinical settings.
Developing a CNN to identify the location of a lesion in an uncropped image and to tolerate nonideal lighting conditions could be useful in some clinical scenarios. Finally, while visual inspection is one of the most important diagnostic tool for identifying vascular anomalies, the overall context is needed to make a clinical decision.

Given the prevalence of vascular anomalies, computer-aided diagnosis has the potential to improve health care outcomes and reduce the cost associated with delayed or incorrect referrals. It may also have particular benefit in resource-limited regions, where tertiary expertise is unavailable but smartphones are increasingly ubiquitous. For future studies, we envision that computer-aided diagnosis of vascular anomalies could not only augment the capabilities of primary care physicians, but also guide specialists in treatment of these conditions. For example, a similar classifier could inform clinicians about outcomes related to infantile hemangioma and predict response to propranolol treatment.

%% file: methods-1.tex
\section*{Methods}

\subsection{Datasets.} This study was approved by the Seattle Children's Institutional Review Board. All data was de-identified in accordance with HIPAA guidelines. Our dataset is composed of clinical data from Seattle Children's from 2002-2018, as well as the dermatology repositories, DermIS\cite{dermis}, DermNet\cite{dermnet}, DermNetNZ\cite{dermnetnz}, DermQuest\cite{dermquest} and the ISIC dermoscopic archive\cite{isic}. The images from Seattle Children's Hospital consist of hemangiomas (infantile and congenital), pyogenic granuloma {(lobular hemangioma)}, venous and glomuvenous malformations, capillary malformations, Sturge-Weber syndrome, spider angioma and lymphatic malformations. The images from the online  repositories consist of pyogenic granuloma, atopic dermatitis, nevus, spider angioma, milia, impetigo, molluscum and tinea.

\subsection{Training algorithm.} To train our dataset, we first removed the final 1000-node softmax layer of the InceptionV3 neural network, we then fine-tune the classifier with our own layers (Supplementary Fig.~\ref{fig:arch}) using the Keras framework. We add a 256-node fully-connected layer, with a ReLu activation, followed by Dropout regularization with a rate of 0.6, and finally a 6-way or 12-way softmax layer, depending on the taxonomy of vascular anomalies being classified. We used the RMSProp optimizer with a learning rate of $1e^{-5}$ and a rho value of 0.9. We used the sklearn library for calculating performance measures including AUC and $F_1$ score.

\subsection{t-SNE algorithm.} The t-SNE plot was generated using  an implementation of Barnes-Hut t-SNE \cite{van2013barnes,pythontsne} using a perplexity value of five, the algorithm was run for 1,000 iterations.

\subsection{Saliency maps.} The saliency maps were generated with an implementation of integrated gradients \cite{sundararajan2017axiomatic,deepviz}. The output is smoothed using the SmoothGrad\cite{smilkov2017smoothgrad} algorithm to produce a sharper map.

\subsection{Run-time analysis.} We timed an implementation of the CNN running in real-time on an iPhone 7. The CNN was ported to the iOS platform using Apple's Core ML tools library which converts the CNN to an iPhone readable format.

\subsection{Data availability statement.} All data necessary for interpreting the manuscript have been included. The datasets used in the current study are not publicly available but may be available from the corresponding authors on reasonable request and with permission of Seattle Children's hospital and the University of Washington. Images from the online repositories were obtained from DermIS\cite{dermis}, DermNet\cite{dermnet}, DermNetNZ\cite{dermnetnz}, DermQuest\cite{dermquest} and the ISIC dermoscopic archive\cite{isic}.

\subsection{Use of human subjects.} All human subjects were practicing  pediatricians and took our tests under informed consent. This study was approved as exempt by the Seattle Children's Institutional Review Board.

%% file: acks-1.tex
\noindent {\bf Acknowledgments.}  The authors thank Jacob Sunshine, and John Thickstun for feedback on the manuscript. The authors also thank Allegro Pediatrics--Issaquah Highlands Group for participation and
Eden Palmer for photography and vascular anomaly photographic archive.

\noindent {\bf Author contributions.} JC designed the algorithms and conducted the analysis with technical supervision by SG;  JC, SR and SG wrote the manuscript; RB and JP edited the manuscript; RB and JP recruited the pediatricians for the study; JP provided the data used in the analysis. SR conceptualized the study.

\noindent {\bf Competing interest statement.} JC, SR, RB and SG have equity stakes in Edus Health, Inc., which is not related to the technology presented in this manuscript. SG is a co-founder of Jeeva Wireless, Inc. and Sound Life Sciences, Inc. RB is a consultant for SpiWay, LLC and a co-founder of EigenHealth, Inc. 

%% file: ms.bbl
\begin{thebibliography}{10}
\expandafter\ifx\csname url\endcsname\relax
  \def\url#1{\texttt{#1}}\fi
\expandafter\ifx\csname urlprefix\endcsname\relax\def\urlprefix{URL }\fi
\providecommand{\bibinfo}[2]{#2}
\providecommand{\eprint}[2][]{\url{#2}}

\bibitem{nosher2014vascular}
\bibinfo{author}{Nosher, J.~L.}, \bibinfo{author}{Murillo, P.~G.},
  \bibinfo{author}{Liszewski, M.}, \bibinfo{author}{Gendel, V.} \&
  \bibinfo{author}{Gribbin, C.~E.}
\newblock \bibinfo{title}{Vascular anomalies: a pictorial review of
  nomenclature, diagnosis and treatment}.
\newblock \emph{\bibinfo{journal}{World journal of radiology}}
  \textbf{\bibinfo{volume}{6}}, \bibinfo{pages}{677} (\bibinfo{year}{2014}).

\bibitem{greene2008risk}
\bibinfo{author}{Greene, A.~K.} \emph{et~al.}
\newblock \bibinfo{title}{Risk of vascular anomalies with down syndrome}.
\newblock \emph{\bibinfo{journal}{Pediatrics}} \textbf{\bibinfo{volume}{121}},
  \bibinfo{pages}{e135--e140} (\bibinfo{year}{2008}).

\bibitem{juern2010nevus}
\bibinfo{author}{Juern, A.~M.}, \bibinfo{author}{Glick, Z.~R.},
  \bibinfo{author}{Drolet, B.~A.} \& \bibinfo{author}{Frieden, I.~J.}
\newblock \bibinfo{title}{Nevus simplex: a reconsideration of nomenclature,
  sites of involvement, and disease associations}.
\newblock \emph{\bibinfo{journal}{Journal of the American Academy of
  Dermatology}} \textbf{\bibinfo{volume}{63}}, \bibinfo{pages}{805--814}
  (\bibinfo{year}{2010}).

\bibitem{haggstrom2006prospective}
\bibinfo{author}{Haggstrom, A.~N.} \emph{et~al.}
\newblock \bibinfo{title}{Prospective study of infantile hemangiomas: clinical
  characteristics predicting complications and treatment}.
\newblock \emph{\bibinfo{journal}{Pediatrics}} \textbf{\bibinfo{volume}{118}},
  \bibinfo{pages}{882--887} (\bibinfo{year}{2006}).

\bibitem{lee2018vascular}
\bibinfo{author}{Lee, J.~W.} \& \bibinfo{author}{Chung, H.~Y.}
\newblock \bibinfo{title}{Vascular anomalies of the head and neck: current
  overview}.
\newblock \emph{\bibinfo{journal}{Archives of craniofacial surgery}}
  \textbf{\bibinfo{volume}{19}}, \bibinfo{pages}{243} (\bibinfo{year}{2018}).

\bibitem{greene2011vascular}
\bibinfo{author}{Greene, A.~K.}, \bibinfo{author}{Liu, A.~S.},
  \bibinfo{author}{Mulliken, J.~B.}, \bibinfo{author}{Chalache, K.} \&
  \bibinfo{author}{Fishman, S.~J.}
\newblock \bibinfo{title}{Vascular anomalies in 5621 patients: guidelines for
  referral}.
\newblock \emph{\bibinfo{journal}{Journal of pediatric surgery}}
  \textbf{\bibinfo{volume}{46}}, \bibinfo{pages}{1784--1789}
  (\bibinfo{year}{2011}).

\bibitem{levin2013room}
\bibinfo{author}{Levin, D.~E.} \emph{et~al.}
\newblock \bibinfo{title}{Room for improvement: Patterns of referral
  misdiagnosis to a vascular anomalies center}.
\newblock \emph{\bibinfo{journal}{Open Journal of Pediatrics}}
  \textbf{\bibinfo{volume}{3}}, \bibinfo{pages}{331} (\bibinfo{year}{2013}).

\bibitem{macfie2008diagnosis}
\bibinfo{author}{MacFie, C.~C.} \& \bibinfo{author}{Jeffery, S.~L.}
\newblock \bibinfo{title}{Diagnosis of vascular skin lesions in children: an
  audit and review}.
\newblock \emph{\bibinfo{journal}{Pediatric dermatology}}
  \textbf{\bibinfo{volume}{25}}, \bibinfo{pages}{7--12} (\bibinfo{year}{2008}).

\bibitem{krizhevsky2012imagenet}
\bibinfo{author}{Krizhevsky, A.}, \bibinfo{author}{Sutskever, I.} \&
  \bibinfo{author}{Hinton, G.~E.}
\newblock \bibinfo{title}{Imagenet classification with deep convolutional
  neural networks}.
\newblock In \emph{\bibinfo{booktitle}{Advances in neural information
  processing systems}}, \bibinfo{pages}{1097--1105} (\bibinfo{year}{2012}).

\bibitem{nixon2012feature}
\bibinfo{author}{Nixon, M.} \& \bibinfo{author}{Aguado, A.~S.}
\newblock \emph{\bibinfo{title}{Feature extraction and image processing for
  computer vision}} (\bibinfo{publisher}{Academic Press},
  \bibinfo{year}{2012}).

\bibitem{esteva2017dermatologist}
\bibinfo{author}{Esteva, A.} \emph{et~al.}
\newblock \bibinfo{title}{Dermatologist-level classification of skin cancer
  with deep neural networks}.
\newblock \emph{\bibinfo{journal}{Nature}} \textbf{\bibinfo{volume}{542}},
  \bibinfo{pages}{115} (\bibinfo{year}{2017}).

\bibitem{rajpurkar2017chexnet}
\bibinfo{author}{Rajpurkar, P.} \emph{et~al.}
\newblock \bibinfo{title}{Chexnet: Radiologist-level pneumonia detection on
  chest x-rays with deep learning}.
\newblock \emph{\bibinfo{journal}{arXiv preprint arXiv:1711.05225}}
  (\bibinfo{year}{2017}).

\bibitem{gulshan2016development}
\bibinfo{author}{Gulshan, V.} \emph{et~al.}
\newblock \bibinfo{title}{Development and validation of a deep learning
  algorithm for detection of diabetic retinopathy in retinal fundus
  photographs}.
\newblock \emph{\bibinfo{journal}{Jama}} \textbf{\bibinfo{volume}{316}},
  \bibinfo{pages}{2402--2410} (\bibinfo{year}{2016}).

\bibitem{poplin2018prediction}
\bibinfo{author}{Poplin, R.} \emph{et~al.}
\newblock \bibinfo{title}{Prediction of cardiovascular risk factors from
  retinal fundus photographs via deep learning}.
\newblock \emph{\bibinfo{journal}{Nature Biomedical Engineering}}
  \textbf{\bibinfo{volume}{2}}, \bibinfo{pages}{158} (\bibinfo{year}{2018}).

\bibitem{pan2010survey}
\bibinfo{author}{Pan, S.~J.} \& \bibinfo{author}{Yang, Q.}
\newblock \bibinfo{title}{A survey on transfer learning}.
\newblock \emph{\bibinfo{journal}{IEEE Transactions on knowledge and data
  engineering}} \textbf{\bibinfo{volume}{22}}, \bibinfo{pages}{1345--1359}
  (\bibinfo{year}{2010}).

\bibitem{szegedy2016rethinking}
\bibinfo{author}{Szegedy, C.}, \bibinfo{author}{Vanhoucke, V.},
  \bibinfo{author}{Ioffe, S.}, \bibinfo{author}{Shlens, J.} \&
  \bibinfo{author}{Wojna, Z.}
\newblock \bibinfo{title}{Rethinking the inception architecture for computer
  vision}.
\newblock In \emph{\bibinfo{booktitle}{Proceedings of the IEEE conference on
  computer vision and pattern recognition}}, \bibinfo{pages}{2818--2826}
  (\bibinfo{year}{2016}).

\bibitem{russakovsky2015imagenet}
\bibinfo{author}{Russakovsky, O.} \emph{et~al.}
\newblock \bibinfo{title}{Imagenet large scale visual recognition challenge}.
\newblock \emph{\bibinfo{journal}{International journal of computer vision}}
  \textbf{\bibinfo{volume}{115}}, \bibinfo{pages}{211--252}
  (\bibinfo{year}{2015}).

\bibitem{kohavi1995study}
\bibinfo{author}{Kohavi, R.} \emph{et~al.}
\newblock \bibinfo{title}{A study of cross-validation and bootstrap for
  accuracy estimation and model selection} \textbf{\bibinfo{volume}{14}},
  \bibinfo{pages}{1137--1145} (\bibinfo{year}{1995}).

\bibitem{cirecsan2012multi}
\bibinfo{author}{Cire{\c{s}}an, D.}, \bibinfo{author}{Meier, U.} \&
  \bibinfo{author}{Schmidhuber, J.}
\newblock \bibinfo{title}{Multi-column deep neural networks for image
  classification}.
\newblock \emph{\bibinfo{journal}{arXiv preprint arXiv:1202.2745}}
  (\bibinfo{year}{2012}).

\bibitem{simard2003best}
\bibinfo{author}{Simard, P.~Y.}, \bibinfo{author}{Steinkraus, D.},
  \bibinfo{author}{Platt, J.~C.} \emph{et~al.}
\newblock \bibinfo{title}{Best practices for convolutional neural networks
  applied to visual document analysis.} .

\bibitem{deng2009imagenet}
\bibinfo{author}{Deng, J.} \emph{et~al.}
\newblock \bibinfo{title}{Imagenet: A large-scale hierarchical image database}.
\newblock In \emph{\bibinfo{booktitle}{2009 IEEE conference on computer vision
  and pattern recognition}}, \bibinfo{pages}{248--255}
  (\bibinfo{organization}{Ieee}, \bibinfo{year}{2009}).

\bibitem{sundararajan2017axiomatic}
\bibinfo{author}{Sundararajan, M.}, \bibinfo{author}{Taly, A.} \&
  \bibinfo{author}{Yan, Q.}
\newblock \bibinfo{title}{Axiomatic attribution for deep networks}.
\newblock In \emph{\bibinfo{booktitle}{Proceedings of the 34th International
  Conference on Machine Learning-Volume 70}}, \bibinfo{pages}{3319--3328}
  (\bibinfo{organization}{JMLR. org}, \bibinfo{year}{2017}).

\bibitem{dermis}
\bibinfo{title}{Derm{IS}.net}.
\newblock
  \bibinfo{howpublished}{https://www.dermis.net/dermisroot/en/home/index.htm}
  (\bibinfo{year}{2019}).

\bibitem{dermnet}
\bibinfo{title}{Dermnet}.
\newblock \bibinfo{howpublished}{www.dermnet.com/} (\bibinfo{year}{2019}).

\bibitem{dermnetnz}
\bibinfo{title}{Dermnet {NZ}}.
\newblock \bibinfo{howpublished}{https://www.dermnetnz.org/}
  (\bibinfo{year}{2019}).

\bibitem{dermquest}
\bibinfo{title}{Derm{Q}uest}.
\newblock \bibinfo{howpublished}{http://dermquest.com} (\bibinfo{year}{2019}).

\bibitem{isic}
\bibinfo{title}{{ISIC}}.
\newblock \bibinfo{howpublished}{https://www.isic-archive.com}
  (\bibinfo{year}{2019}).

\bibitem{van2013barnes}
\bibinfo{author}{Van Der~Maaten, L.}
\newblock \bibinfo{title}{{B}arnes-{H}ut-{SNE}}.
\newblock \emph{\bibinfo{journal}{arXiv preprint arXiv:1301.3342}}
  (\bibinfo{year}{2013}).

\bibitem{pythontsne}
\bibinfo{title}{Python-{TSNE}}.
\newblock \bibinfo{howpublished}{https://github.com/danielfrg/tsne}
  (\bibinfo{year}{2019}).

\bibitem{deepviz}
\bibinfo{title}{deep-viz-keras}.
\newblock \bibinfo{howpublished}{https://github.com/experiencor/deep-viz-keras}
  (\bibinfo{year}{2019}).

\bibitem{smilkov2017smoothgrad}
\bibinfo{author}{Smilkov, D.}, \bibinfo{author}{Thorat, N.},
  \bibinfo{author}{Kim, B.}, \bibinfo{author}{Vi{\'e}gas, F.} \&
  \bibinfo{author}{Wattenberg, M.}
\newblock \bibinfo{title}{Smoothgrad: removing noise by adding noise}.
\newblock \emph{\bibinfo{journal}{arXiv preprint arXiv:1706.03825}}
  (\bibinfo{year}{2017}).

\end{thebibliography}
